\documentclass[aps,prl,twocolumn,groupedaddress,showpacs]{revtex4}
\usepackage{epsfig}
\usepackage{amsmath}
\usepackage{amsfonts}
\def\bE{\mathbf E}
\def\bH{\mathbf H}
\def\bJ{\mathbf J}

\def\bx{\mathbf x}

\def\b0{\mathbf 0}

\begin{document}

\title{Observation of negative impulse velocity in free space}
\author{Neil V. Budko}
\affiliation{Laboratory of Electromagnetic Research, Faculty of Electrical
Engineering, Mathematics and Computer Science,
Delft University of Technology,
Mekelweg 4, 2628 CD Delft, The Netherlands}
\email{n.v.budko@tudelft.nl}

\date{\today}

\begin{abstract}
Since the 1983 definition of the speed of light in vacuum as a fundamental constant with the exact value of 299792458 m/s 
the question remains as to what apart from the wavefront travels at that speed. It is commonly assumed that the entire 
wave-packet or an impulse of the electromagnetic radiation in free space does. Here it is shown, both theoretically and experimentally, 
that there exists a region close to the source, where, while the wave-front travels at the speed of light, the individual impulses comprising 
the body of the wave-packet appear to slow down and even go backwards in time. This three-dimensional near-field late-time effect may 
also explain some of the free-space superluminal measurements.
\end{abstract}


\maketitle
Around the year 1983 it has been decided to link the etalon of length to the speed of light in vacuum \cite{Giacomo}. 
This has effectively put an end to the measurements of the ``actual'' velocity of light which is now defined to have an exact numerical value. 
However, there remains a seemingly simple question: what exactly travels at 299792458 m/s? It is more or less universally agreed that the 
wavefront of the electromagnetic pulse in vacuum does. Although,
there is no formal proof of this fact. It is also often assumed that the electromagnetic waveform, which follows the wavefront, also travels
at that speed in free space. This assumption is essential for such technologies as the radar ranging, travel-time tomography, and information transfer which all
rely on an extremum or a slope of an impulse, or the envelope of a wave-packet \cite{Skolnik,Valle}. Yet, these practically useful waveform features do not 
always travel at the speed of light. 

It is a well-known fact that dispersive media may drastically change the shape of the envelope, so that the group velocity significantly deviates from
the velocity of light in the bulk medium \cite{LandauLifshitz,Diener,Hau,Wang,Bajcsy}. Recently, it has been reported that ``localized microwaves'' travel
at what appears to be a superluminal velocity in free space \cite{Mugnai}. Although, the proposed theory and the accuracy of the measurements were subsequently 
disputed, there are no fundamental constraints prohibiting such behavior \cite{Wynne,Porras}. Here we look at the problem from a more general point of view, without
specific reference to the Bessel-like X-beams. It turns out that the changing impulse velocity in free-space is an inherent feature of the three-dimensional 
near-field radiation. Instead of accelerating, though, the individual impulses are actually ``slowing down''. This slow-light behavior is spatially-dependent and 
gradually disapears as the distance to the source increases. The latter phenomenon leads to the faster-than-light results for the impulse velocity.
Here it is also shown that in the immediate neighborhood of the source the impulses inside a wave-packet may even travel back in time, 
which could be naively interpreted as an anti-causal behavior.

According to the Maxwell equations, which govern the propagation of the electromagnetic field, the source of the field is the electric current density 
$\bJ$:
 \begin{align}
 \label{eq:Maxwell}
 \begin{split}
 -\nabla\times \bH(\bx,t)+\varepsilon_{0}\partial_{t}\bE(\bx,t)&=-\bJ(\bx,t),
 \\
 \nabla\times\bE(\bx,t)+\mu_{0}\partial_{t}\bH(\bx,t)&=0.
 \end{split}
 \end{align}
Mathematically, causality enters here in the form of an assumption that the changes in the current {\it cause} the changes in the field. 
In particular, it is typically assumed that the field is zero everywhere before the current is switched on. Once the current density of the source 
is given as a function of space and time, and the surrounding medium is a free space, the Maxwell equations can be solved analytically, 
thus providing an explicit radiation formula for the fields everywhere outside the source at all times. In subscript notation to denote the 
Cartesian components of the electric field strength and the current density, using Einstein's summation convention, this important formula 
may be written as follows (see e.g. \cite{deHoop}):
 \begin{align}
 \label{eq:Radiation}
 \begin{split}
 E_{k}&(\bx,t)=
 \\
 &\int\limits_{\bx'\in D}\frac{1}{\varepsilon_{0}4\pi\vert\bx-\bx'\vert^{3}}
 \left(3\theta_{k}\theta_{n}-\delta_{kn}\right)\int\limits_{-\infty}^{t_{\rm R}}J_{n}(\bx',t')\,{\rm d}t'\,{\rm d}\bx'
 \\
 &+\int\limits_{\bx'\in D}\frac{1}{c_{0}\varepsilon_{0}4\pi\vert\bx-\bx'\vert^{2}}
 \left(3\theta_{k}\theta_{n}-\delta_{kn}\right)J_{n}(\bx',t_{\rm R})\,{\rm d}\bx'
 \\
 &+\int\limits_{\bx'\in D}\frac{1}{c_{0}^{2}\varepsilon_{0}4\pi\vert\bx-\bx'\vert}
 \left(\theta_{k}\theta_{n}-\delta_{kn}\right)\partial_{t}J_{n}(\bx',t_{\rm R})\,{\rm d}\bx'.
 \end{split}
 \end{align}
Where  $\theta_{n}=(x_{n}-x_{n}')/\vert\bx-\bx'\vert$ represents a unit vector pointing from the location $\bx'$  inside the finite spatial 
domain  $D$ occupied by the source towards an observer situated at $\bx$; $\delta_{kn}$  is the Kronecker delta; 
and $t_{\rm R}=t-\vert\bx-\bx'\vert/c_{0}$ is the so-called retarded time. It is this last simple expression for the retarded time, 
which one is tempted to apply to deduce the speed of light from the observed time-delay for a travelling electromagnetic impulse. 
Yet, the retarded time enters the three terms of the above radiation formula in three different ways. The terms correspond to the {\it near-}, 
{\it intermediate-}, and {\it far-field} zones, in accordance with the relative dominance of their spatial decay factors. Sufficiently away from the 
source, the last term dominates. Therefore, in the far-field zone the temporal dependence of the field will look like a somewhat 
distorted (differentiated) but otherwise just a shifted-in-time and reduced-in-magnitude copy of the current. Hence, having done 
measurements at two known locations, one further away from the source than the other, we can apply the retarded-time formula 
to some obvious feature of the wave-packet, say a particular extremum, and recover an approximate value of the speed of light. 
Approximate, because the influence of the near-field term, although small, is not exactly zero even in the far-field zone (we shall come 
back to this at the end). Closer to the source, however, the time-domain dynamics of the field is much more complicated. 
Here we focus our attention on the rapidly decaying near-field term, which contains the time-integral of the current density 
from the switch-on moment up to the retarded time. Obviously, this integral will have more influence at later times, i.e. 
inside the wave-packet, leaving the early-time wave-front practically unchanged. In particular, nothing but causal retardation will happen 
to the initial moment when impulse rises above zero. This explains why the wave-front travels at the speed of light. 
To better understand the effect of the near-field term on the overall time-dependence, consider a current, which (as a function of time) 
has a well-defined maximum at a certain instant  $t_{\rm M}$, and let us observe the field at some fixed spatial location. 
If this location is in the immediate neighbourhood of the source, then the maximum (or minimum) will be shifted forward in time 
and observed at some new (later) instant  $t_{\rm M}'\ge t_{M}$. There is nothing ``superluminal'' about this fact, as it is the 
usual {\it local} phase shift between the current and the voltage in an AC circuit, which could be deduced already from the first of the
Maxwell equations (\ref{eq:Maxwell}). Hence, the electric field immediately outside the source will, in general, be shifted forward (later) 
in time with respect to the current density. Once we are in the air and move further away from the source, the influence of the 
time-integral and the near-field term on the overall time-dependence will rapidly diminish in inverse proportion to the cube of the 
distance (\ref{eq:Radiation}). In fact, the influence of the near-field term diminishes so rapidly that the field impulses inside a 
wave-packet may appear to be moving backwards in time as the distance increases. 

To illustrate this effect let us consider a simple point-dipole model of the source, where the spatial integrals in the radiation formula 
are evaluated analytically \cite{deHoop,Jackson}. The resulting formula is especially simple for the direction of observation orthogonal 
to the direction of the dipole moment (see Figure~1),
\begin{figure}[t]
\vspace{0.3cm}
\epsfig{file=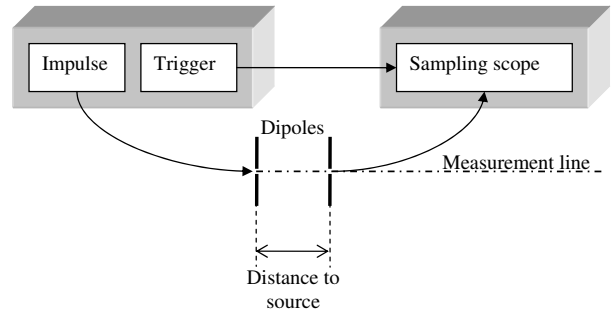,width=8cm}
\caption{The near-field measurement setup with two identical vertically polarized dipole antennas placed at 
a varying distance along the measurement line.}
\end{figure}
since along the measurement line the electric field vector is parallel to the current in the source. 
We take the excitation current to be a finite sinusoidal wave-packet with 4~GHz central frequency and compute the waveforms at increasing 
distances from the source spanning the spatial interval from 10~mm to 100~mm. The results are presented in Figure 2, where the top 
part gives all the waveforms as a single image. The horizontal axis is time ( in metres), the vertical axis is the distance to the source along 
the measurement line, and the amplitude of the electric field strength is encoded in the colour. The near-field zone corresponds to the bottom 
area of the image. 
\begin{figure}[t]
\epsfig{file=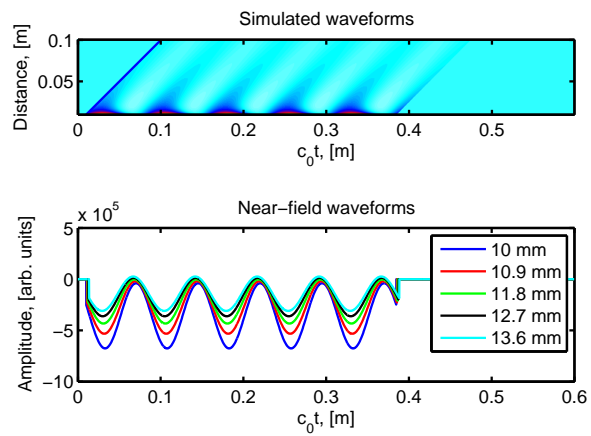,width=8.5cm}
\caption{Simulation of the near-field measurement setup with two non-interacting point-dipole antennas fed by a sinusoidal wave-packet. 
Top: space-time wave-packet dynamics; lower part of this image corresponds to the near-field zone. 
Bottom: the first five waveforms; the outer edges of the wave-packet shift rightwards – normal (light cone) behaviour; 
the inner impulses shift leftwards – negative impulse velocity.}
\end{figure}
A signal travelling at the speed of light should follow (be parallel to) the so-called light cone, which looks as a rising 45~deg 
line for convenience shown in our image. We see that the leftmost part of the wave-packet – its wave-front – does indeed follow the light cone. 
However, the impulses, which comprise the body of the wave-packet, at small distances make an initial bend to the left and begin to follow the 
light cone only somewhat further from the source. In the bottom plot of Figure~2 we can see the details of the first five near-field waveforms, 
where it is clearly visible that, while the edges of the wave-packet shift to the right, the inner impulses shift to the left, i.e. travel backwards 
in time. Paradoxically, an observer in the immediate vicinity of the source will receive the sequence of impulses somewhat later than another 
observer slightly further away!

The difficulty with performing a near-field experiment verifying our prediction is in the presence of strong coupling between the source and 
the receiver antennas, which changes the impedance and thus the shape of the waveform in addition to any other changes one is trying to detect. 
In the far-field the source-receiver coupling may sometimes be seen as a series of barely visible multiple scattering events. In the near-field such 
obvious identification is no longer possible. In reality, the whole phenomenon of impulses moving back in time could be completely masked or 
even negated by the coupling. In addition the finite size of the dipoles is expected to introduce further deviations from the point-dipole model.

The actual experiment was performed using a pair of unbalanced 31~mm long copper wire dipoles. The source antenna was fed by an 
approximately Gaussian impulse, with 71.4~ps FWHM duration (Picosecond Pulse Labs impulse generator, model~3500). The internal trigger of the 
generator is used to trigger the Agilent Infiniium DCA-J~86100C mainframe with a 20~GHz sampling module (Agilent~86112A), which recorded and 
averaged the waveform received by the second dipole. Both antennas naturally deform the original Gaussian waveform, so that the received 
signal looks like a sequence of impulses, required for the verification of our hypothesis.The results are presented in Figure~3 in the form identical to 
the simulations of Figure~2. 
\begin{figure}[t]
\epsfig{file=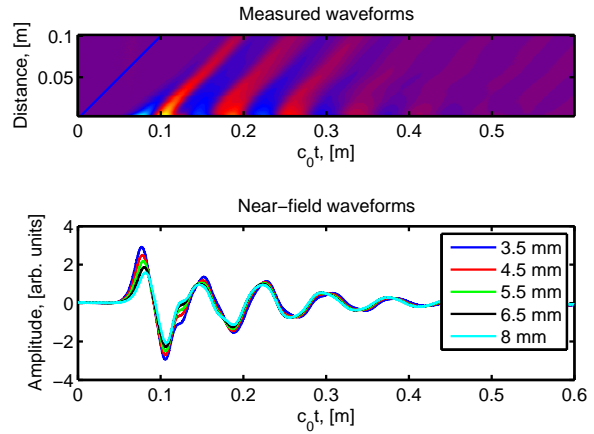,width=8.5cm}
\caption{Experimental observation of the free-space negative impulse velocity. Top: space-time wave-packet dynamics; 
the leftwards bend in the lower part of the image indicates the negative impulse velocity in the near-field. 
Bottom: details of the first five waveforms with secondary peaks travelling back in time.}
\end{figure}
Despite the mentioned inadequacy of the point-dipole model, the similarity with simulations is obvious, as we see both 
the leftward bend in the bottom part of the top space-time image and the back-in-time motion of the secondary peaks of the first five near-field 
waveforms shown in the bottom plot of Figure~3.

The present phenomenon may also explain some of the earlier apparently superluminal observations based on the time-delay and distance 
measurements \cite{Mugnai}. Consider two observers (A and B) recording waveforms at different distances from the source. The influence of the near-field 
term will be stronger for the nearest of the two observers (A) and weaker for the other (B). Not only in the near-field zone, but also further away, the impulses will keep 
shifting backwards in time, although this shift is now relatively small compared to the overall forward motion of the wave-packet. In any case, (A) 
will see the impulses shifted {\it more forward}, if compared to the same impulses as seen by (B). Hence, the time-delay between identical impulses 
at (A) and (B) would appear to be shrinking, resulting in an apparently superluminal velocity of propagation approaching the speed of light 
asymptotically from above as the distance from the source increases. 

Finally a few words have to be said about the terminology and the physical interpretation of the observed phenomenon. Although the terms 
``group'' and ``phase'' velocities come to mind, in the title and throughout the text the term ``impulse velocity'' is used instead. 
The reason is that both the group and the phase velocities are well-defined for plane waves only, whereas here we have a purely three-dimensional 
phenomenon. Indeed, the near-field term appears in the radiation formula only after summation (inverse Fourier transform) of the $k$-domain solution 
of the Maxwell equations over all possible wave-vectors \cite{deHoop,Jackson}. All this demonstrates that, apart from the wavefront, a definite velocity can 
hardly be associated with any particular feature of the electromagnetic waveform. In such circumstances, it makes sense to go back to the 
original concept of the speed of light as a mere numerical constant which together with either $\varepsilon_{0}$ or $\mu_{0}$  
determines the mathematical form of the Maxwell equations. On the practical side, the present results show that algorithms relying on the 
impulse velocity require significant correction beyond the simple travel-time schemes when applied in the near-field zone.
The effect of the intermediate-field term on the overall time-dependence should be further investigated along similar lines. However, the corresponding 
experiments may turn out to be more difficult, due to the anticipated smallness of the effect.


\end{document}